\documentclass{PoS}

\usepackage{graphicx,color,amsmath}

\def\beq{\begin{eqnarray}}
\def\eeq{\end{eqnarray}}

\title{Update of the 2HDM-III with a four-zero texture in the Yukawa matrices and phenomenology of the charged Higgs Boson}

\ShortTitle{Update of the 2HDM-III with a four-zero texture in the Yukawa matrices}

\author{ \speaker {J. Hern\' andez-S\' anchez} %
\thanks{  
J. H.-S. thanks the University of Southampton and the Rutherford Appleton Laboratory for hospitality
during his visit to the NExT Institute where  part of this work was carried out. 
J. H.-S, A. R.-S and R. N.-P acknowledge the financial 
support of SNI, CONACYT, SEP-PROMEP and VIEP-BUAP. 
We all thank A.G. Akeroyd for innumerable constructive and useful discussions.} \\
Fac. de Cs. de la
Electr\'onica, Benem\'erita Universidad Aut\'onoma de Puebla, Apdo. Postal 542, 72570 Puebla, Puebla, M\'exico and Dual C-P Institute of High Energy Physics, M\'exico. \\
E-mail: \email{jaimeh@ece.buap.mx} }

\author{S. Moretti %
 \thanks{S.M. is supported in part through the NExT Institute.} \\
School of Physics and Astronomy, University of Southampton, Highfield, Southampton SO17 1BJ, United Kingdom, and Particle Physics Department, Rutherford Appleton Laboratory, Chilton, Didcot, Oxon OX11 0QX, United Kingdom \\ 
E-mail: \email{s.moretti@soton.ac.uk} }

\author{R. Noriega-Papaqui \\
\'Area Acad\'emica de Matem\'aticas y F\'{\i}sica, Universidad Aut\'onoma del Estado
de Hidalgo, Carr. Pachuca-Tulancingo Km. 4.5, C.P. 42184, Pachuca, Hgo. and Dual C-P Institute of High Energy Physics, M\'exico. \\
E-mail: \email{rnoriega@uaeh.edu.mx} }

\author{A. Rosado \\
Instituto de
F\'{\i}sica, BUAP, Apdo. Postal J-48, C.P. 72570 Puebla, Pue.,
M\'exico.
E-mail: \email{rosado@ifuap.buap.mx} }

\abstract{
We update the flavor-violating constraints on the charged Higgs sector  of the   2-Higgs Doublet Model Type-III (2HDM-III), using a four-zero texture in the Yukawa matrices. 
We give a generic Lagrangian of the 2HDM-III. In order to show the relevance of the off-diagonal terms
of such a texture, we utilize the main  constraints from $B$-physics, $\mu -e$ universality in $\tau$ decays and  the radiative decay $Z\to b \bar{b}$ presented recently in arXiv:1212.6818 [hep-ph].
In particular,  we show that the $H^- c \bar{b}$   coupling can be very large 
and very different  with respect to 
2HDMs with a flavor discrete symmetry (i.e., ${\mathcal{Z}}_2$). We also discuss the possible enhancements  of the vertices $H^\pm W^\mp V$ ($V=Z, \gamma$)  that  arise at one-loop level.}
 
\FullConference{Prospects for Charged Higgs Discovery at Colliders\\
                 October 8-11, 2012\\
                 Uppsala University Sweden}
 
\begin{document}

\section{Introduction}

Once discovered the new particle at LHC \cite{latest_Higgs}, which is very compatible with the neural Higgs Boson of the  Standard Model
(SM) \cite{stanmod}. The LHC aims to search for physics beyond SM. In particular the flavor physics could be a scope of interesting, where
 the main problem is to control the presence of Flavor
Changing Neutral Currents (FCNCs). 
In the most general version of a 2HDM, the fermionic couplings of the neutral scalars are non-diagonal in flavor and, therefore, generate unwanted FCNC phenomena \cite{Branco:2011iw}.  The simplest and most common approach is to impose a $\mathcal{Z}_2$ symmetry forbidding all non-diagonal  terms in flavor space
in the Lagrangian \cite{Glashow:1976nt}. 

In particular, we focus here on the
version where the Yukawa couplings depend on the
hierarchy of masses. This version is the one where the mass matrix has a four-zero
texture form \cite{fritzsch} forcing the non-diagonal Yukawa couplings to be proportional to the geometric mean of the two fermion masses, $g_{ij}\propto \sqrt{m_i m_j} \chi_{ij}$ \cite{Cheng:1987rs,
DiazCruz:2004pj}. This matrix is based on the phenomenological
observation that the off-diagonal elements must be small in order to
dim the interactions that violate flavor, as experimental results
show. 
 In this work, we  will discuss the increase in sensitivity to the Branching Ratio (BR) for $H^\pm \to cb,  W^\pm \gamma, W^\pm Z$ and 
to the fermionic couplings of a $H^\pm$ in the 2HDM-III scenario. 

%%%%%%%%%%%%%%%%%%%%%%%%%%%%%%%%%%%%%%%%%%%%%%%%%%%%%
\section{The  Higgs-Yukawa sector  of the 2HDM-III}
%%%%%%%%%%%%%%%%%%%%%%%%%%%%%%%%%%%%%%%%%%%%%%%%%%%%

The 2HDM includes two Higgs scalar doublets of hypercharge $+1$:
$\Phi^\dag_1=(\phi^-_1,\phi_1^{0*})$ and
$\Phi^\dag_2=(\phi^-_2,\phi_2^{0*})$. The most general $SU(2)_L \times U(1)_Y $
invariant  scalar potential  can be written as~\cite{Gunion:2002zf}
\begin{eqnarray}
V(\Phi_1,\Phi_2)&=&\mu^2_1(\Phi_1^\dag
\Phi_1)+\mu^2_2(\Phi^\dag_2\Phi_2)-\left(\mu^2_{12}(\Phi^\dag_1\Phi_2)+{\rm
H.c.}\right) + \frac{1}{2}
\lambda_1(\Phi^\dag_1\Phi_1)^2 \\ \nonumber 
&& +\frac{1}{2} \lambda_2(\Phi^\dag_2\Phi_2)^2+\lambda_3(\Phi_1^\dag
\Phi_1)(\Phi^\dag_2\, \Phi_2)
+\lambda_4(\Phi^\dag_1\Phi_2)(\Phi^\dag_2\Phi_1) \\ \nonumber 
&& +
\left(\frac{1}{2} \lambda_5(\Phi^\dag_1\Phi_2)^2+\left(\lambda_6(\Phi_1^\dag
\Phi_1)+\lambda_7(\Phi^\dag_2\Phi_2)\right)(\Phi_1^\dag \Phi_2)+
{\rm H.c.}\right), \label{potential}
\end{eqnarray}
where all parameters are assumed to be real\footnote{The  $\mu^2_{12}$, $\lambda_5$, $\lambda_6$ and $\lambda_7$ parameters are complex in general, but we will assume that they are real for simplicity.}. 
When a
specific four-zero texture is implemented as a flavor symmetry in the Yukawa sector, discrete
symmetries in the Higgs potential are not needed.
Thence, one must keep the terms proportional to $\lambda_6$ and $\lambda_7$.  These parameters play an important 
role in one-loop processes, where self-interactions
of Higgs bosons could be relevant \cite{HernandezSanchez:2011fq}. 

%%%%%%%%%%%%%%%%%%%%%%%%%%%%%%%%%%%%%%%%%%%%%%%%%%%%%%%%%%%%%%%%%
%\section{The Yukawa sector in the 2HDM-III with a four-zero texture}
%%%%%%%%%%%%%%%%%%%%%%%%%%%%%%%%%%%%%%%%%%%%%%%%%%%%%%%%%%%%%%%%%%
Otherwise, in order to derive the interactions of the type Higgs-fermion-fermion, the
Yukawa Lagrangian is written as follows: 
{\small 
\beq 
 {\cal{L}}_{Y}  = -\Bigg(
Y^{u}_1\bar{Q}_L {\tilde \Phi_{1}} u_{R} +
                   Y^{u}_2 \bar{Q}_L {\tilde \Phi_{2}} u_{R} +
Y^{d}_1\bar{Q}_L \Phi_{1} d_{R}  
 + Y^{d}_2 \bar{Q}_L\Phi_{2}d_{R} +Y^{{l}}_{1}\bar{L_{L}}\Phi_{1}l_{R} +Y^{{l}}_{2}\bar{L_{L}}\Phi_{2}l_{R} \Bigg),
\label{lag-f} 
\eeq }
\noindent where $\Phi_{1,2}=(\phi^+_{1,2},
\phi^0_{1,2})^T$ refer to the two Higgs doublets, ${\tilde
\Phi_{1,2}}=i \sigma_{2}\Phi_{1,2}^* $.
%%%%%%
After spontaneous EW Symmetry Breaking (EWSB), one can derive the
fermion mass matrices from eq. (\ref{lag-f}), namely: 
$ M_f= \frac{1}{\sqrt{2}}(v_{1}Y_{1}^{f}+v_{2}Y_{2}^{f})$, $f = u$, $d$, $l$.
Assuming that both Yukawa matrices $Y^f_1$ and $Y^f_2$ have the
four-texture form and are Hermitian \cite{DiazCruz:2004pj}.
 The diagonalization is
performed in the following way:
$ \bar{M}_f = V_{fL}^{\dagger}M_{f}V_{fR}$. Then, 
$\bar{M}_f=\frac{1}{\sqrt{2}}(v_{1}\tilde{Y}_{1}^{f}+v_{2}
\tilde{Y}_{2}^{f})$, 
where $\tilde{Y}_{i}^{f}=V_{fL}^{\dagger}Y_{i}^{f}V_{fR}$.
One can derive a better approximation for the product
$V_q\, Y^{q}_n \, V_q^\dagger$, expressing the
rotated matrix $\tilde {Y}^q_n$ as:
\begin{eqnarray}
\left[ \tilde{Y}_n^{q} \right]_{ij}
= \frac{\sqrt{m^q_i m^q_j}}{v} \, \left[\tilde{\chi}_{n}^q \right]_{ij}
=\frac{\sqrt{m^q_i m^q_j}}{v}\,\left[\chi_{n}^q \right]_{ij}  \, e^{i \vartheta^q_{ij}},
\label{cheng-sher}
\end{eqnarray}
\noindent
where the $\chi$'s are unknown dimensionless parameters of the model.  
Following the recent analysis of \cite{HernandezSanchez:2012eg}, we can obtain the generic expression
for the interactions of  Higgs bosons with the fermions:
{\small
\begin{eqnarray}
{\cal L}^{\bar{f}_i f_j \phi}  & = &
-\left\{\frac{\sqrt2}{v}\overline{u}_i
\left(m_{d_j} X_{ij} {P}_R+m_{u_i} Y_{ij} {P}_L\right)d_j \,H^+
+\frac{\sqrt2m_{{l}_j} }{v} Z_{ij}\overline{\nu_L^{}}{l}_R^{}H^+
+{H.c.}\right\} \nonumber \\
& &-  
\frac{1}{v} \bigg\{ \bar{f}_i m_{f_i} h_{ij}^f  f_j h^0 + \bar{f}_i m_{f_i} H_{ij}^f  f_j H^0 - i \bar{f}_i m_{f_i} A_{ij}^f  f_j \gamma_5 A^0\bigg\},
\label{lagrangian-f}
\end{eqnarray}
where $\phi_{ij}^f$ ($\phi=h$, $H$, $A$), $X_{ij}$, $Y_{ij}$ and $Z_{i j}$ are defined as:
\begin{eqnarray}\label{hHA}
\phi_{ij}^f & = & \xi_\phi^f \delta_{ij} + G(\xi_\phi^f,X), \, \, \, \phi= h, H, A,   \nonumber \\
X_{i j} & = &   \sum^3_{l=1}  (V_{\rm CKM})_{il} \bigg[ X \, \frac{m_{d_{l}}}{m_{d_j}} \, \delta_{lj}
-\frac{f(X)}{\sqrt{2} }  \,\sqrt{\frac{m_{d_l}}{ m_{d_j} }} \, \tilde{\chi}^d_{lj}  \bigg],
\nonumber \\
Y_{i j} & = &  \sum^3_{l=1}  \bigg[ Y  \, \delta_{il}
  -\frac{f(Y)}{\sqrt{2} }  \,\sqrt{\frac{ m_{u_l}}{m_{u_i}} } \, \tilde{\chi}^u_{il}  \bigg]  (V_{\rm CKM})_{lj},
  \label{Xij}
\nonumber \\
Z_{i j}^{l}& = &   \bigg[Z \, \frac{m_{{l}_{i}}}{m_{{l}_j}} \, 
\delta_{ij} -\frac{f(Z)}{\sqrt{2} }  \,\sqrt{\frac{m_{{l}_i}}{m_{{l}_j}}  }
\, \tilde{\chi}^{l}_{ij}  \bigg].
\label{Zij}
\end{eqnarray} }
 where $G(\xi_\phi^f,X)$ can be obtained from  \cite{HernandezSanchez:2012eg} and the parameters $\xi_\phi^f$, $X$, $Y$ and $Z$ are given in the Table \ref{couplings}. When the parameters $\chi_{ij}^f=0$,   one recovers the Yukawa interactions given in Refs.~\cite{Grossman:1994jb, Akeroyd:2012yg,Aoki:2009ha}.  
 As was pointed in   \cite{HernandezSanchez:2012eg}, we suggest that this Lagrangian could represent a 
Multi-Higgs Doublet Model (MHDM)
or an Aligned 2HDM (A2HDM) with additional flavor physics in the Yukawa matrices as well as the possibility of FCNCs at tree level. 
 %% %%%%%%%%%%%%%%%%%%%%%%%%%%%%
%Tab. 1%
%%%%%%%%%%%%%%%%%%%%%%%%%%%%%%
{\small
\begin{table}
\begin{center}
\begin{tabular}{|c|c|c|c|c|c|c|c|c|c|}
\hline
 2HDM-III& $X$ &  $Y$ &  $Z$ & $\xi^u_h $  & $\xi^d_h $ & $\xi^l_{h} $  & $\xi^u_H $  & $\xi^d_H $ & $\xi^{l}_H $\\ \hline
2HDM-I-like
&  $-\cot\beta$ & $\cot\beta$ & $-\cot\beta$ & $c_\alpha/s_\beta$ & $c_\alpha/s_\beta$ & $c_\alpha/s_\beta$  
& $s_\alpha/s_\beta$ & $s_\alpha/s_\beta$ & $s_\alpha/s_\beta$\\
2HDM-II-like
& $\tan\beta$ & $\cot\beta$ & $\tan\beta$ & $c_\alpha/s_\beta$ & $-s_\alpha/c_\beta$ & $-s_\alpha/c_\beta$ 
& $s_\alpha/s_\beta$ & $c_\alpha/c_\beta$ & $c_\alpha/c_\beta$\\
2HDM-X-like
& $-\cot\beta$ & $\cot\beta$ & $\tan\beta$ &  $c_\alpha/s_\beta$ & $c_\alpha/s_\beta$ & $-s_\alpha/c_\beta$  
& $s_\alpha/s_\beta$ & $s_\alpha/s_\beta$ & $c_\alpha/c_\beta$\\
2HDM-Y-like
& $\tan\beta$ & $\cot\beta$ & $-\cot\beta$ & $c_\alpha/s_\beta$ & $-s_\alpha/c_\beta$ & $c_\alpha/s_\beta$   
& $s_\alpha/s_\beta$ & $c_\alpha/c_\beta$ & $s_\alpha/s_\beta$\\ 
\hline
\end{tabular}  
\end{center} 
\caption{ Parameters $\xi_\phi^f$, $X$, $Y$ and $Z$  defined in the  Yukawa interactions of eq.   (3.4) for four versions of the 2HDM-III with a four-zero texture. Here $s_\alpha = \sin \alpha $, $ c_\alpha = \cos \alpha $, 
$s_\beta = \sin \beta $ and $ c_\beta = \cos \beta $. }
\label{couplings}
\end{table} }

%%%%%%%%%%%%%%%%%%%%%%%%%%%%%%%%%%%%%%%%%%%%%
\section{Flavor constraints on  the  2HDM-III with a four-zero Yukawa texture}
%%%%%%%%%%%%%%%%%%%%%%%%%%%%%%%%%%%%%%%%%%%%%%%%%%

According to the recent analysis of flavor constraints on the 2HDM-III with a four-zero texture reported in \cite{HernandezSanchez:2012eg},
 we now enumerate those for the new physics  parameters $\chi_{ij}^f$  that come from a four-zero Yukawa texture.  
 %%%%%%%%%%%%%
\begin{itemize}
\item $\mu - e$ universality in $\tau$ decays
\end{itemize}
%%%%%%%%%%%%%%%%%
The $\tau$ decays  into $ \mu \bar{\nu}_{\mu} \nu_{\tau}$ and $ e \bar{\nu}_{e} \nu_{\tau}$
produce important constraints onto charged Higgs boson states coupling to 
  leptons, through the requirement of  
 $\mu - e$ universality. 
We can get the   constraint:
$\frac{|Z_{22} Z_{33}|}{m_{H^\pm}^2} \leq 0.16 $ GeV$^{-1}$   (95$\% $ CL).
We obtain that the parameter space more favored is when $0.8 \leq |\chi_{ij}|\leq 2$,  for  $0.5 \leq Z \leq 100$, $X\leq 80$ and $m_{H^\pm} \geq 100$ GeV.
%%%%%%%%%
\begin{itemize}
\item Leptonic meson decays  $B\to\tau\nu$, $D \to \mu \nu$,  $D_s \to \mu \nu, \tau \nu$ and  semileptonic decays  $B\to D \tau \nu$
\end{itemize}
%%%%%%%%%%%%%
Given the leptonic decays $B\to\tau\nu$, $D \to \mu \nu$,  $D_s \to \mu \nu, \tau \nu$ and the semileptonic ones  $B\to D \tau \nu$,
we show in the Table \ref{bounds-chi23}  the constraints for the off-diagonal terms $\chi_{23}^{u,d}$  of the Yukawa texture. We assume $0.1\leq \chi_{22}^{l} =  \chi_{33}^{l} \leq 1.5  $ as well as $  \chi_{22}^d =   \chi_{22}^u =1$
and take  80 GeV $\leq m_{H^\pm} \leq 160$ GeV. 
%% %%%%%%%%%%%%%%%%%%%%%%%%%%%%
%Tab. 2%
%%%%%%%%%%%%%%%%%%%%%%%%%%%%%%
\begin{table}
\begin{center}
\begin{tabular}{|c||c|c|c|c|}
\hline
  2HDM-III's & $\chi_{23}^d (B\to \tau \nu)$   &   $\chi_{23}^u(D_s\to l \nu)$  
&   $\chi_{23}^u(B\to D \tau \nu)$ &  $\chi_{23}^u$ (combination)   \\ \hline
2HDM-I-like & (-0.35,-0.15) or &  (-1.5,0.9)   & (-0.05,0.45)   &  (-0.05,0.45)  \\   &   (0,0.15)  & & &   \\ \hline
 2HDM-II-like  &  (-0.35,-0.2) or   & (-2,27)    &   (-9.6,1.2) 
&  (-2,1.2)     \\  &     (0,0.2) & & &   \\ \hline
 2HDM-X-like &   (-7,2)  &  (-4,14) & (-3.8,0.47)   &    (-4,0.47)  \\ &&&& \\ \hline
2HDM-Y-like  &  (-1.8,-1.2) or  & (-40,50)  &   (-16,50)  &   (-16,50)  
\\ &   (-0.2,0.6)  & & &  \\ \hline
\end{tabular} 
\end{center}
\caption{ Constraints from   $B \to D \tau \nu$, $D_s \to \tau \nu, \mu \nu$ and $B \to \tau \nu$
decays. We show the allowed intervals  for off-diagonal terms $\chi_{23}^{u,d}$. }
\label{bounds-chi23}
\end{table}
%%%%%%%%%%%%%%%%%%%%%
\begin{itemize}
\item $B \to X_s \gamma$ decays
\end{itemize}
Using  values  for the charged Higgs boson mass in the interval
80 GeV $\leq m_{H^\pm} \leq 300$ GeV, we can establish the following constraints:
$\bigg|\frac{Y_{33} Y_{32}^*}{V_{tb} V_{ts}}\bigg|  < 0.25$,
$  -1.7<Re\bigg[\frac{X_{33} Y_{32}^*}{V_{tb} V_{ts}}\bigg]< 0.7$.
 One can then extract the bounds $-0.75\leq \chi_{23}^u \leq -0.15$ for $\chi_{33}^u=1$ and  
$0.4\leq \chi_{23}^u \leq 0.9$ for $\chi_{33}^u=-1$, both when $Y<<1$. Assuming the allowed interval for   
$ \chi_{23}^d $ from $B\to \tau\nu$ and 
  $\chi_{33}^u=1=$ $\chi_{33}^d=1$.  We can, e.g., obtain $ \chi_{23}^u \in (-0.55,-0.48)$ for the case $X=20$ and $ Y=0.1 $. Another interesting scenario for the 2HDM-III is the 2HDM-X-like one, where the allowed region is larger than in other scenarios, with $ \chi_{23}^u \in (-2.2, 0.45)$  and 
  $ \chi_{23}^d \in (-7, -2)$.
%%%%%%%%%%%%%%%%%%%%%%%%%
\begin{itemize}
\item  $B^0-\bar{B}^0$ mixing\end{itemize}
Considering the  areas allowed by the measured $\Delta M_{B_d}$ value
within a $2 \sigma$ error,  when we have a light charged Higgs (80 GeV $\leq m_{H^\pm} \leq 200$ GeV) one can  extract the limit
$\bigg|\frac{Y_{33}^* Y_{31}}{V_{tb} V_{td}}\bigg|  \leq 0.25$,
which is consistent with the bounds obtained from $B\to X_s \gamma$.

In summary, the diagonal terms are such that $\chi_{kk}^f \sim 1$ and the off-diagonal terms  such that $|\chi_{ij}^f| \leq 0.5$, which establish a parameter space region allowed by these constraints. 

%%%%%%%%%%%%%%%%%%%%%%%%%%%%%%%%%%%%
\section{The dominance of the BR$(H^\pm \to cb)$}
%%%%%%%%%%%%%%%%%%%%%%%%%%%%%%%%%%%%%
A distinctive signal of a $H^\pm$ state from the 2HDM-III for $m_{H^\pm}< m_t-m_b$ would be a 
sizable BR for $H^\pm \to cb$. 
For $m_{H^\pm} < m_t-m_b$, the scenario of $|X|>> |Y|,|Z|$ in a 2HDM-III 
gives rise to a ``leptophobic'' $H^\pm$ with 
BR$(H^\pm\to cs)$+BR$(H^\pm\to cb) \sim 100\%$, as the BR($H^\pm \to \tau\nu$)
is negligible ($<<1\%$). 
Conversely, one can see that the configuration $Y >>$ $X$,$ Z$ (this imply that $Y_{ij} >>$ $X_{ij}$,$ Z_{ij}$, see eqs. (\ref{Xij})
 is very interesting, because the decay $H^+ \to c \bar{b}$ is now dominant. 
In order to show this situation, we calculate the dominant terms $m_c Y_{23}$, $m _c Y_{22}$ of the width $\Gamma(H^+ \to c \bar{b}, c\bar{s})$, respectively, which are given by:
$m_c Y_{cb}  =  m_c Y_{23}  = V_{cb}m_c \bigg(Y - \frac{f(Y)}{\sqrt{2}} \chi_{22}^u \bigg)- V_{tb} \frac{f(Y)}{\sqrt{2}} \sqrt{m_t m_c} 
\chi_{23}^u$ and 
$m_c Y_{cs} =  m_c Y_{22} = V_{cs}m_c \bigg(Y - \frac{f(Y)}{\sqrt{2}} \chi_{22}^u \bigg)- V_{ts} \frac{f(Y)}{\sqrt{2}} \sqrt{m_t m_c} 
\chi_{23}^u$.
As $Y$ is large and $f(Y) = \sqrt{1+Y^2} \sim Y $, then the term $\bigg( Y - \frac{f(Y)}{\sqrt{2}} \chi_{22}^u \bigg)$ could be absent or small, when $\chi_{ij} = O(1)$. Besides,  the last term is very large because $ \propto\sqrt{m_t m_c}$, given that $m_t=173$ GeV, so that in the
end this is the dominant contribution. Therefore, 
we can compute the ratio of two dominant decays, namely,  BR$(H^\pm\to cb)$ and BR$(H^\pm\to cs)$, which is given as follows:
\begin{equation}
R_{sb}=\frac{{\rm BR}(H^\pm\to cb)}{{\rm BR}(H^\pm\to cs)}\sim \frac{|V_{tb}|^2 }{|V_{ts}|^2 }.
\label{Rsb-1}
\end{equation}
In this case  BR$(H^\pm\to cb) \sim 100\% $ (for $m_{H^\pm} < m_t-m_b$, of course) so that to verify
this prediction would really be the hallmark signal  of the 2HDM-III. Therefore, we can see 
 that the non-diagonal term $\chi_{23}^u$  cannot be omitted and 
this is an important result signalling new physics even beyond the standards 2HDMs.  
Another case is when $X >>$ $Y$,$ Z$, here, we get that the dominants terms are
$\propto m_b X_{23}$, $m _s X_{22}$, 
$m_b X_{cb} =  m_b X_{23}  = V_{cb}m_b \bigg(X - \frac{f(X)}{\sqrt{2}} \chi_{33}^d \bigg)- V_{cs} \frac{f(X)}{\sqrt{2}} \sqrt{m_b m_s} 
\chi_{23}^d$ and 
$m_s X_{cs}  =  m_s X_{22} = V_{cs}m_s \bigg(X - \frac{f(X)}{\sqrt{2}} \chi_{22}^d \bigg)- V_{ts} \frac{f(X)}{\sqrt{2}} \sqrt{m_b m_s} 
\chi_{23}^d$ .
 In this case there are two possibilities. If $\chi= O(1)$ and positive then $\bigg(X - \frac{f(X)}{\sqrt{2}} \chi_{33}^d \bigg)$ is small and
 \beq
  R_{sb}\sim \frac{|V_{cs}|^2 }{|V_{cb}|^2 }. 
  \label{Rsb-2}
  \eeq
  Here, the BR$(H^\pm \to cb)$ becomes large, again, this case too could be another exotic scenario of the 2HDM-III. 
 The other possibility is when  $\chi= O(1)$ and negative, then 
  \beq
  R_{sb}\sim \frac{m^2_b |V_{cb}|^2 }{m^2_s |V_{cs}|^2 }, 
  \label{Rsb-3}
  \eeq
  which is very similar to the cases studied recently in \cite{Akeroyd:2012yg}. In summary, one can see two possibilities to study the 
  BR$(H^\pm \to cb)$: firstly, the scenarios given in eqs. (\ref{Rsb-1}) and (\ref{Rsb-2}),
which are peculiar to the 2HDM-III; secondly, the scenario very close to the MHDM/A2HDM expressed in eq. (\ref{Rsb-3}).

In a similar spirit, one can study other interesting channels, both in  decay and production and both at tree and one-loop
level \cite{DiazCruz:2004pj,  HernandezSanchez:2011fq}. For instance, in processes at one-loop level is possible to combine two effects in the model: firstly, the term $\bigg( Y - \frac{f(Y)}{\sqrt{2}} \chi_{22}^u \bigg)$ (or $X$ for $d$) could be absent or small, when $\chi_{ij} \sim 1$;
secondly, the self-interaction $H^+ H^- \phi^0 $ can be large because the Higgs potential parameters $\lambda_{6,7}$ are sizable. Then, one can typically
get a BR$(H^+ \to W^+ \gamma ) \sim 10^{-4}, \, \,10 ^{-3}$ and a BR$(H^+ \to W^+ Z ) \sim 10^{-3}$, $10 ^{-2} $.

\section{Conclusions}

In our model (2HDM-III), 
BR($H^\pm \to cb$) could be as large as $90\%$. 
 Along the lines indicated by previous literature, 
in the context of the 2HDM-III with a four-zero Yukawa texture,
we would conclude by suggesting that a dedicated search for $t\to H^\pm b$ and $H^\pm \to cb$ 
or $cb\to H^\pm$ would
probe values of the fermionic couplings of $H^\pm$ that are compliant with current searches.
The outlook is therefore clear. Depending on the search channel, both the Tevatron (possibly) and the LHC (certainly)
have the potential to constrain or else discover the 2HDHM-III supplemented by a four-zero
Yukawa texture including non-vanishing off-diagonal terms in the Yukawa matrices.

\end{document}